# HIGHLY EFFICIENT CRYSTAL DEFLECTOR FOR CHANNELING EXTRACTION OF A PROTON BEAM FROM ACCELERATORS

V. Guidi, C. Malagù, G. Martinelli, M. Stefancich, D. Vincenzi, *Univ. Ferrara, INFM-INFN, Italy*;
V.M. Biryukov, Yu.A. Chesnokov, V.I. Kotov, *IHEP Protvino, Russia*; W. Scandale, *CERN, Switzerland*

*Abstract*
The design and performance of a novel crystal deflector for proton beams are reported. A silicon crystal was used to channel and extract 70 GeV protons from the U-70 accelerator in Protvino with an efficiency of (85.3±2.8)%, as measured for a beam of ~$10^{12}$ protons directed towards crystals of ~2 mm length in spills of ~2 s duration. Experimental data agree with the theoretically predicted Monte Carlo results for channeling. The technique allows one to manufacture a very short deflector along the beam direction (2 mm). Consequently, multiple encounters of circulating particles with the crystal are possible with little probability of multiple scattering and nuclear interactions per encounter. Thus, drastic increase in efficiency for particle extraction out of the accelerator was attained. We show the characteristics of the crystal-deflector and the technology behind it. Such an achievement is important in devising a more efficient use of the U-70 accelerator and provides crucial support for implementing crystal-assisted slow extraction and collimation in other machines, such as the Tevatron, RHIC, the AGS, the SNS, COSY, and the LHC.

## INTRODUCTION

The use of bent crystals for beam extraction in accelerators is under development at several laboratories. [1,2,3]. The advantages of this method are the ease of its realisation, feasibility of its simultaneous work with collider regime or with internal targets, and the absence of intensity pulsations because no resonant blow-up of the beam is needed to direct the beam onto the crystal for extraction. The crystal has a minimal "septum width", and is very convenient even for application in a beam-loss localisation system as a coherent scatterer [4].

A collaboration of researchers working at the 70-GeV accelerator of IHEP has recently achieved a substantial progress in the parameters of crystal-assisted beam deflection: an extraction efficiency larger than 85% has been obtained up to such an high intensity as $10^{12}$ protons [4]. A major feature of such reaching was the usage of very short crystals for extraction; the crystals were selected to the minimal value foreseen by the physics of channeling [5,6]. Thereby, the circulating particles encountered the crystal many times and suffered negligible divergence at each pass due to reduced scattering and nuclear interactions in the crystal. Multiple passage of particles allowed the protons to be eventually channeled by the crystal planes, leading to the experimentally recorded high efficiency.

Fig. 1 shows the theoretical prediction [7] made for the extraction efficiency at 70-GeV proton accelerator, and the figures measured in the campaign of measurements through 1997-2001 with crystals of different size and design. The trend is clear from both theory and experiments: the shorter the crystal, the higher the efficiency. The theoretical plot indicates the optimal crystal length for the conditions of the 70-GeV experiment, i.e. slightly less than 1 mm. Indeed such a short crystal deflector is very difficult to obtain. In this paper, we discuss the design and manufacturing of the crystal deflectors that led to the successful execution of experiments.

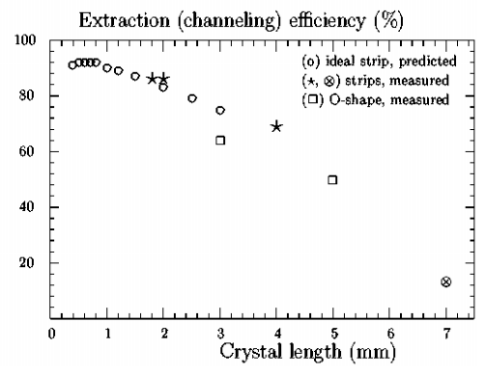

Fig.1 Crystal extraction efficiency as measured and as predicted for 70-GeV protons.

## CRYSTAL BENDING SYSTEM

The first designs of short crystals appeared in literature in 1998 [3]. It consisted of an "O-shaped" deflector cut from a monolithic piece of oriented silicon. Here the bending was obtained by compressing the crystal at its middle part. Indeed, this method is not suitable for an efficient deflection with a crystal shorter than 5 mm. Any attempt to reduce this length resulted in poor bending efficiency as it is clear in Fig. 1 with the "O-shaped" 3-mm long sample.

Therefore, a new design to produce suitably short sample is needed. A possibility is the use of the anisotropic properties of a crystal lattice. From the theory of elasticity it is known (see e.g.[8]) that bending a crystal plate in the longitudinal direction causes some "anticlastic bending" or twists appear in the orthogonal direction. In that case, a crystal plate obtains the shape of a saddle, barrel, or a pure cylinder depending on the concrete anisotropic properties of the material. The surface equation for the crystal plate in the ideal case of the method of moments discussed in [9] is:

$$y = \frac{1}{2R_{II}} z^2 - kx^2$$

Where $y$ is the direction of the incident beam, $x$ and $z$ are the longitudinal and transversal coordinates, and $k$ is a coefficient depending on the concrete anisotropic properties of the material (see Fig. 2). For $k>0$ this is the equation of a hyperbolical paraboloid (saddle), $k<0$ gives the equation of an elliptical paraboloid (barrel), and $k=0$ is for a parabolic cylinder. It turns out that the first case applies to Si (111), therefore the crystal plate obtains the shape of a saddle as sketched in Fig.2. The orthogonal bending of a narrow crystal plate was used for beam deflection.

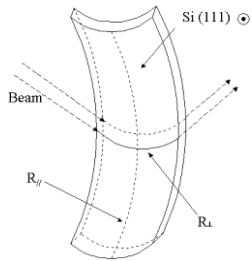

**Fig. 2 Scheme of the bent crystal plate.**

In our case, a crystal plate of silicon may be shaped like a saddle for the purpose (Fig. 2). The technical advantages of such a deflecting system are that it may be easily made shorter along the beam, has no straight sections in bending, and needs no additional material around the "legs".

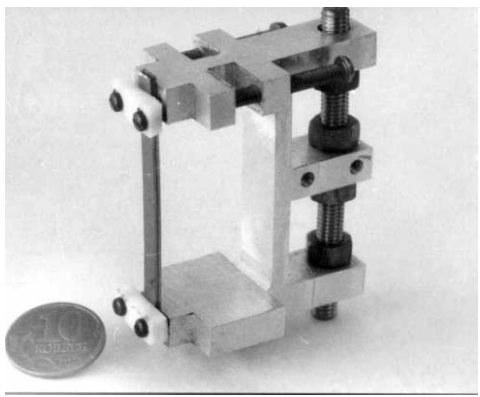

**Fig. 3 Photo of crystal bending device (on the left is a narrow crystal strip).**

The crystals were manufactured at the Semiconductors and Sensors Laboratory of the University of Ferrara as narrow strips, about 2 mm along the beam. The metal holder showed in Fig. 3 imparts the right curvature to the samples. A strip of monocrystalline silicon was bent in the longitudinal direction by an angle of about 100 mrad in the orthogonal direction: as a consequence the strips were bent in the orthogonal direction of about 1 mrad. The quality of the strip bending was preliminary checked by the laser system described in Ref. [8]. Then the crystals were tested directly in the experiments on high-energy accelerator.

## SAMPLE'S PREPARATION

The starting material consisted of prime-grade, (111) oriented, 525-µm-thick silicon wafers. The crystals were sliced to form 0.5×50×2 mm$^3$ (thickness, height, and length, respectively) by means of a mechanical dicing saw. The length of 2 mm was chosen to be exposed to steer the protons in the collider and was determined according to the previous considerations. The (111) crystalline direction was intended for alignment toward the radial direction in the accelerator.

A crucial methodology for the achievement of high-performance crystal extractors was the chemical treatment of the samples as explained in next section. The preparation of the sample herewith described is suitable to remove the defects induced by the saw at the beam entry and exit facets of the silicon crystals. Indeed, mechanical slicing of the samples induces a large amount of scratches, dislocations, line defects and anomalies that would reduce the overall channelling efficiency of the crystal. According to the experience gained through previous runs at the accelerator, it came out that a 30 µm superficial layer was ineffective for channeling. We attributed this effect to the presence of in-depth lattice imperfections induced during the slicing.

Thus, we attempted the removal of such a layer by a sequence of room-temperature chemical treatments to the surfaces [10]. To avoid an unwanted sample thinning, the largest surfaces were protected by Apiezon wax while the others were left uncoated. The prime treatment was a wet planar etching (HF, HNO$_3$ and CH$_3$COOH (3:5:3)) and the timing was such to obtain an etching depth of about 30 µm. The samples were prepared and treated in clean-room environment (class 100) to avoid material contamination, which would locally modify the etch rates and the finishing of the etch ground.

Preliminary, the first process stages were aimed at the removal of pollutants such as greasy or metallic compounds. The wafers were degreased in trichloro-ethylene, acetone and then isopropanol. A two-stage cleaning procedure was, then, carried out to remove organic and metallic impurities from the surface of the wafers. The samples were cleaned in solution of water, hydrogen peroxide and ammonium hydroxide (5:1:1) at 75 °C for 10 min. After a short dipping in diluted HF (10% in weight) the wafers were washed in water, hydrogen peroxide and hydrochloric acid (4:1:1) at 75 °C for 10 min. The specimens were cut from the coated wafer through a diamond-blade saw avoiding any alignment with major crystalline axes.

Finally, the entry and exit facets were ready for the prime chemical etching to remove the mechanical damages induced by the blade. The wax coating was eventually removed according to standard procedure and the resulting samples were once more subject to a complete cleaning process.

## TESTS WITH HIGH-ENERGY PROTONS

The new crystal-deflectors were first tested in an external beam of 70 GeV protons. Fig. 4 shows a schematic of the experimental set-up. The incident proton beam was monitored by scintillation counters S1, S2. The crystal holder in the beam line may be tilted through a goniometer around the axis orthogonal to the proton beam and to the (111) orientation of the crystal. The rotation of the goniometer can be imparted as steps of 5 μrad each. The optimum crystal orientation was obtained by tilting the sample through a goniometer to determine the maximum counting rate of the remote scintillation counters S3, S4.

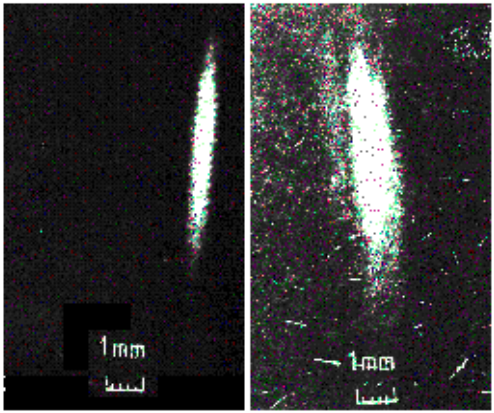

**Fig.4 Image of the beam deflected with chemically polished crystal (left) and mechanically treated (right).**

The precise measurement of the profiles of the bent beam has been done with nuclear photo-emulsions placed outside the vacuum chamber, 7 m far from the crystal.

Fig. 4 shows the results of the tests for the chemically polished deflectors compared to the deflectors having mechanical treatment of end faces. As seen from the picture, the profile of the beam bent by a chemically polished crystal is more uniform and sharp. Its width corresponds to the crystal thickness once the beam divergence within the critical angle of channeling (equal to 20 μrad for 70 GeV protons) has been taken into account. At the same time, the beam bent by a mechanically polished crystal has irregularities corresponding to an angular distortion of the order of 100 μrad.

After testing in an external beam, the crystals were installed in the IHEP U-70 accelerator ring to extract the circulating beam. The crystals with chemically polished faces have shown the best efficiency for beam extraction — up to a value larger than 80% [4]. New crystals can be applied for beam extraction and collimation of beams at accelerators in a broad range of energies as shown in Ref.[4,11]. One of the major applications would be the use in large colliders such as the LHC as a collimation system.

## ACKNOWLEDGEMENTS

This work was supported by INTAS-CERN Grant No. 132-2000, RFBR Grant No. 01-02-16229, and by the "Young Researcher Project" of the University of Ferrara.